\documentclass{easychair}

\usepackage{amsmath}
\usepackage{amsthm}
\usepackage{amssymb}
\usepackage{newlfont}
\usepackage{amsmath}
\usepackage{latexsym}
\usepackage{mathrsfs}
\usepackage{url}
\usepackage{graphics,epsfig,times}
\usepackage{stmaryrd}
\usepackage{xcolor}
\usepackage{graphicx}
\usepackage{setspace}
\usepackage{lscape}
\usepackage{listings}

\newtheorem{example}{Example}

\usepackage[draft,mode=multiuser]{fixme} 

\begin{document}

%
\title{Automating the Generation of Cyber Range Virtual Scenarios with VSDL}

%
\titlerunning{Automating the Generation of CR Virtual Scenarios with VSDL}

\volumeinfo
	{\emph{Journal of Wireless Mobile Networks, Ubiquitous Computing, and Dependable Applications (JoWUA)}}    	
	{13}                         	
	{1}          
	{33}                         	
	{Mar. 2022}
	{\emph{10.22667/JOWUA.2022.03.31.0033}}

%

\author{\\
Gabriele Costa${}^{1}$, Enrico Russo${}^{2}\thanks{Corresponding author: Department of Informatics, Bioengineering, Robotics, and Systems Engineering (DIBRIS), Viale Causa, 13, 16145 Genoa, Italy}$, and Alessandro Armando${}^{2}$\\[1em]
${}^{1}$IMT School for Advanced Studies, Lucca 55100 Italy\\
gabriele.costa@imtlucca.it\\[1em]
${}^{2}$University of Genoa, DIBRIS, Genoa 16145 Italy\\
enrico.russo@unige.it, alessandro.armando@unige.it\\
{\scriptsize Received: October 17, 2022; Accepted: December 4, 2022; Published: December 31, 2022}
}

%
\authorrunning{Costa, Russo and Armando}

\maketitle

%
\begin{abstract}
\noindent
A \emph{cyber range} (CR) is an environment used for training security experts and testing attack and defense tools and procedures.
Usually, a cyber range simulates one or more critical infrastructures that attacking (red) and defending (blue) teams must compromise and protect, respectively.
The infrastructure can be physically assembled, but much more convenient is to rely on the Infrastructure as a Service (IaaS) paradigm.
Although some modern technologies support the IaaS, the design and deployment of scenarios of interest are mostly manual.
As a consequence, it is a common practice to have a cyber range hosting few (sometimes only one), consolidated scenarios.
However, reusing the same scenario may significantly reduce the effectiveness of the training and testing sessions.

In this paper, we propose a framework for automating the definition and deployment of arbitrarily complex cyber range scenarios.
The framework relies on the \emph{virtual scenario description language} (VSDL), i.e., a domain-specific language for defining high-level features of the desired infrastructure while hiding low-level details.
The semantics of VSDL is given in terms of constraints that must be satisfied by the virtual infrastructure.
These constraints are then submitted to an SMT solver to check the satisfiability of the specification.
If satisfiable, the specification gives rise to a model that is automatically converted to a set of deployment scripts to be submitted to the IaaS provider. 
\end{abstract}

\section{Introduction}
\label{sec:intro}

Cyber defense (as well as offense) relies on security experts that must be properly trained through hands-on activities~\cite{Demetrio2019} and equipped with adequate security tools.
This simple fact is boosting the interest of the international community toward creating \emph{cyber ranges}.
In short, a cyber range is an environment where trainees compete or cooperate to achieve specific security goals.
Needless to say, they should interact with a realistic environment that accurately mimics real-world settings.
A common practice is to have a defending, aka \emph{blue}, team and an attacking, aka \emph{red}, team.
For instance, the blue team can be asked to enhance the security of an Information Technology (IT) infrastructure in a limited amount of time.
Afterward, the red team must violate the security of the infrastructure by accessing target data or compromising a certain resource.
Specific security vulnerabilities of interest can also be injected into the original infrastructure to organize aimed sessions.
The operational environment, including networks, hardware, software, and how they behave during the session, is called a \emph{scenario}.

Infrastructure as a Service (IaaS) is a convenient paradigm for defining and deploying the elements of a scenario.
Virtualization technologies can emulate both networks and computational nodes.
For instance, OpenStack~\cite{sefraoui2012openstack} can be used to emulate large-scale, heterogeneous networks of virtual machines.
However, defining virtual scenarios for a cyber range poses several specific issues.
The main limitation is the relatively short lifetime of a scenario.
In principle, a cyber range should permit the definition of a scenario that is used for a session lasting a few hours.
Ideally, the scenarios should not be reused as they might become repetitive and rapidly lose interest.
Thus, the cyber range should provide a mechanism for rapidly generating new scenarios while guaranteeing that they include the desired features.
Unfortunately, the process of defining a scenario using the existing IaaS solutions is nontrivial.
Indeed, IaaS platforms are usually designed for defining and hosting long-term infrastructures.
As a consequence, existing cyber ranges tend to reuse few scenarios.
For instance, the Michigan cyber range~\cite{Merit16webpage} has a single virtual infrastructure of four buildings of a fictional town called Alphaville.
Similarly, the Open Cyber Challenge Platform (OCCP)~\cite{wagner2013designing} includes a limited number of scenarios.\footnote{For the time being, only one scenario is documented, and few others have been announced.}
Instead, the annual NATO Locked Shields initiative~\cite{absil2016lessons} relies on a virtual scenario renewed yearly and only used for two days.
As mentioned above, reusing the same scenario drastically reduces the effectiveness of the training activity and, thus, the usefulness of the cyber range.

\begin{figure*}[t]
\begin{center}
\includegraphics[width=\textwidth]{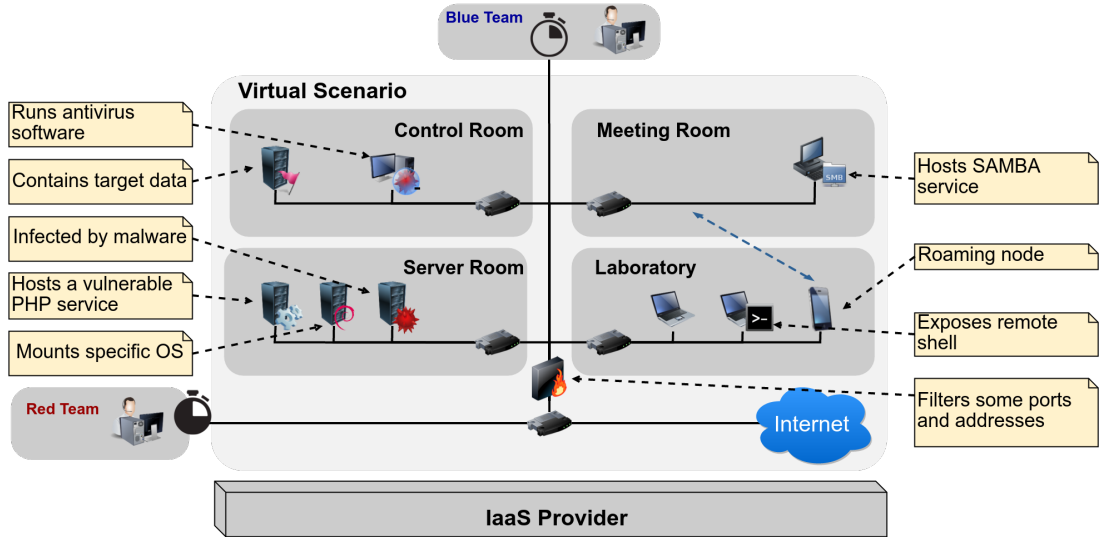} 
\caption{A virtual scenario with annotated requirements.}
\label{fig:virtualscen1}
\end{center}
\end{figure*}

We propose an example to highlight better the structure of a virtual scenario and its features.
We will use it as a working example along the paper to constructively show the steps of our approach.

\begin{example}
\label{ex:base}
Consider the scenario graphically depicted in Figure~\ref{fig:virtualscen1}.
It consists of a network composed of four sub-networks, i.e., Server Room, Laboratory, Meeting Room, and Control Room.
The blue team is deployed within the network perimeter and goes offline after a certain time.
Instead, the red team access is delayed and takes place from outside the perimeter, i.e., from the public network.

The features of the elements appearing in the scenario (written inside note labels) are heterogeneous and partial.
Most of them are straightforward.
For instance, the Laboratory network must consist of two laptops and a mobile phone.
One of the laptops is accessible through a remote shell, and the mobile phone, at a certain time, must migrate to the network Meeting Room.
Notice that no other aspects of these nodes are relevant to the scenario, e.g., the structure of the filesystem of the three devices. 
\end{example}

Reasonably, the features that one wants to specify when defining a scenario belong to the following categories.
\begin{description}
\item[Networking.] What are the existing channels and connections? Are there active firewalls? What rules do they apply?
\item[Hardware.] What are the hardware capabilities of the nodes (CPU speed, disk size)? What is their role in the scenario (e.g., servers, mobile phones, or laptops)?
\item[Software.] What OS runs on a node? What applications and libraries are installed? Is there a software monoculture~\cite{Geer03mono}?
\item[Data.] What information is stored in a node? Does the file system contain any relevant files?
\item[Users and privileges.] Who can access a certain node? What are the privileges of the users over the file system of a node?
\item[Time.] How does the infrastructure change over time? Are there nomadic nodes? Are there failures of nodes or networks?
\end{description}

We claim that domain-specific languages \cite{vanDeursen00dsl} (DSL) are in order.
As a matter of fact, their tailored syntax can precisely describe the desired features of any infrastructure.
Also, the formal semantics of a DSL supports and drives automatic validation, refinement, and implementation processes.  

In this paper, we present a framework for the automatic validation and implementation of virtual scenarios for cyber ranges.
The framework relies on a \emph{virtual scenario description language} (VSDL) for the high-level specification of the scenario properties.
The semantics of a VSDL specification is given in terms of (quantifier-free) linear integer arithmetic (QFLIA, see~\cite{Barrett15smt}) assertions.
Assertions are then processed through a \emph{satisfiability modulo theories} (SMT) solver, which checks whether they admit a model.
The model assigns values to constants and functions, and it is automatically translated into a corresponding virtual scenario.

Our approach provides several advantages over the manual modeling of a scenario.
Among them, the most important ones are the following.
\begin{description}
\item[Verifiability.] Ensuring that a scenario exposes some relevant features, e.g., the presence of a vulnerability, is usually nontrivial. 
Satisfiability checking returns a model with the requested properties.
The model is automatically translated into a scenario script, so avoiding errors that might derive from a manual implementation.
\item[Expressiveness.] VSDL permits the description of a scenario using a rich syntax.
Expressible statements cover many aspects of the scenario, from the high, abstract level to low, concrete details.
Temporal conditions can guard statements so that scenarios evolving in time can also be modeled.
\item[Compositionality.] Existing scenarios can be modified and extended by adding new statements, elements, or even entire infrastructures. 
For instance, a scenario can be created by combining the infrastructures used in other previously defined scenarios.
The verification process guarantees that the composition does not invalidate the required features.
\item[Integration.] The result of the instantiation process is a set of scripts for the deployment engine of an IaaS provider.
As far as they refer to distinct entities, the scripts can be combined with those produced through other channels (e.g., manually written).
\end{description}

A further contribution of this work is the implementation of a working prototype that has been integrated with state-of-the-art technologies like OpenStack, Terraform, and Packer.

\emph{This paper is structured as follows.} 
Section~\ref{sec:back} describes some related works on cyber ranges and virtual infrastructures.
In Section~\ref{sec:arch} we provide an overview of the architecture of our framework, while in Section~\ref{sec:vsdl}, we present the virtual scenario description language and its interpretation.
Finally, Section~\ref{sec:agen} presents the virtual scenario generation procedure, and Section~\ref{sec:conc} concludes the paper.


\section{Related work}
\label{sec:back}

A cyber range hosts one or more virtual infrastructures used for training and testing purposes (see~\cite{Davis13survery} for a survey).
The number of such facilities is rapidly growing and many active projects exist, e.g., see~\cite{AZCWR16cwr,Benzel11deter,Merit16webpage,wagner2013designing,Ferguson14ncr}.

The \emph{NATO Locked Shield}~\cite{absil2016lessons} and the \emph{National Cyber Range} 
\cite{Ferguson14ncr} (NCR) are among the most prominent initiatives.
We already mentioned in the introduction the first one and the related yearly events.
Instead, the NCR relies on a physical infrastructure partially documented in \cite{Rosenstein12ncr}.
However, at the best of our knowledge, more recent proposals tend to avoid bare metal implementation as it is more costly and less flexible.

In general, virtual infrastructures are attracting major interest by both academia and industry.
The main reason is that they decouple the computational elements from the physical infrastructure hosting them.
This favors their re-usability, maintainability, adaptiveness and resilience.

The technologies supporting Infrastructure as a Service (IaaS) are the main candidates for the implementation of a cyber range.
For instance, IBM's \emph{Softlayer}~\cite{Softlayer16webpage} and VMWare's \emph{vCloud} \cite{vCloud16webpage} can be used to deploy a private cloud.
Some environments, e.g., Cisco's \emph{Fog Computing}~\cite{syed2016pattern}, even support mixed cloud-IoT infrastructures. 
Nevertheless, these solutions target long-term infrastructures, while cyber range scenarios can have a very short life.
 
Some authors proposed specification languages for describing virtual infrastructures.
For instance, in~\cite{Ghijsen12towards}, the \emph{Infrastructure and Network Description Language} (INDL) is presented.
There the authors show that INDL is expressive enough to model two virtual infrastructures of interest studied by two EU projects.
Another proposal is the description language VXDL~\cite{Koslovski09vxdl}.
With VXDL one can define the requirements that the infrastructure must satisfy to achieve its goal, e.g., in terms of latency.
These languages can precisely describe a virtual infrastructure.
However, because of their different purpose, they are not adequate for the definition of virtual scenarios for the cyber range whereby structural aspects may not be strictly defined, e.g., disk size of a server or connection bandwidth, while some requirements must be satisfied, e.g., the presence of a piece of vulnerable software on some node.

Network virtualization can be carried out through \emph{Software Defined Networking}~\cite{McKeown08openflow,Kreutz12sdn} (SDN). 
SDN allows for the definition and the centralized management of virtual networks abstracted from the physical layer but does not support the description of computational nodes connected to the virtual networks.

A number of languages for the definition and orchestration of services have been put forward, see, e.g.,~\cite{Bartoletti06types,Costa12modular,Montesi07jolie,Castagna07theory,Barbosa09perspective}.
Some of these frameworks can automatically generate service compositions that satisfy functional or security goals. 
Although web services can be relevant or even central in a scenario, these languages do not provide adequate support to model the infrastructural elements.

The possibility of injecting vulnerabilities is crucial for the scenarios of the cyber range. 
Several applications and systems for testing vulnerability scanners and training security analysts have been released in the last few years.
For instance, \emph{Damn Vulnerable Web Application} (DVWA)~\cite{Dvwa16webpage}, \emph{WebGoat}~\cite{Webgoat16webpage} and \emph{Gruyere}~\cite{Gruyere16webpage} deliberately include vulnerabilities and challenges.
Similar projects target other environments of interest, e.g.,~OSes and mobile apps.
Even though they are relevant, the training with these applications tends to be artificial and repetitive and, thus, partially incompatible with the requirements of a cyber range.  

A further crucial aspect is the number of available resources.
As a matter of fact, a virtual infrastructure is executed by a physical platform.
In principle, given a scenario, there is no guarantee that the physical environment has enough computational resources to execute it effectively.
Some existing cyber ranges~\cite{Merit16webpage,wagner2013designing} avoid this check by using a few scenarios that have been extensively tested.
However, it is important to notice that, due to their strategic role, national authorities tend not to divulge the internals of their cyber ranges.


\section{Architecture and Workflow}
\label{sec:arch}

In this section, we describe our approach in terms of the used technologies and how we compose them into a unified workflow.

\subsection{Involved technologies}

We integrate our proposal with state-of-the-art technologies supporting the creation and management of virtual infrastructures.
Such integration guarantees that our approach can be readily applied to the real world.

\paragraph{OpenStack.}

OpenStack~\cite{Corradi14openstack} is a platform for the execution of private and public clouds.
Many providers, e.g. IBM, VMWare, Cisco, Citrix, etc.\footnote{see \url{https://www.openstack.org/marketplace/drivers/} for a complete list.}, joined the initiative by integrating the OpenStack API into their products.
The OpenStack framework consists of a collection of core services dedicated to all the aspects of a virtual infrastructure.
Moreover, it provides APIs and a dashboard application that an administrator can use to create, modify and monitor the existing infrastructures.

\paragraph{Terraform.}
The OpenStack dashboard is designed for manually defining a virtual infrastructure.
Terraform~\cite{Terraform16webpage} provides a convenient way for creating and managing them.
In particular, Terraform relies on a scripting language that one can use to describe the virtual elements, e.g., nodes and networks.
Then, Terraform translates the script content into a sequence of OpenStack API invocations to create the defined objects.
Moreover, a script can be used to update existing elements.
Indeed, Terraform automatically checks whether differences exist between the running infrastructure and the new script and only submits the needed modifications.
Clearly, a Terraform script must precisely describe the elements to be created.

\paragraph{Packer.}
Another task that one might want to automate is the creation and configuration of node images.
This operation requires generating an OS image that must be installed on a node and configuring it with the required software.
Similarly to Terraform, Packer~\cite{Packer16webpage} offers a convenient scripting language for defining and customizing OS images.
A terraform script can exploit one or more images created with Packer for initializing a computational node.

\subsection{Workflow}

\begin{figure*}[t]
\begin{center}
\includegraphics[width=\textwidth]{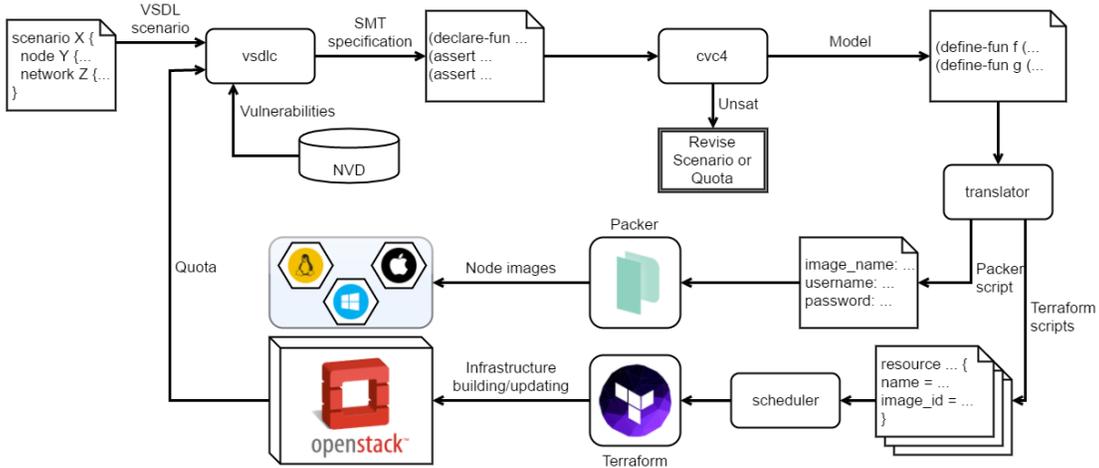} 
\caption{Schematic representation of the VSDL-based workflow.}
\label{fig:virtualscen}
\end{center}
\end{figure*}

Figure~\ref{fig:virtualscen} depicts the abstract workflow of our approach.
We start from a running instance of OpenStack.
Such instances can host one or more virtual infrastructures, created by different administrators, called tenants.
Each tenant has a \emph{quota} of assigned, virtual resources (e.g., virtual CPUs and virtual disk size).
A tenant wanting to create a new scenario writes a VSDL specification as described in Section~\ref{sec:syntax}.
Then, the specification is processed by the VSDL compiler (i.e., \emph{vsdlc}) which also retrieves the quota information from OpenStack.
Moreover, the compiler collects the definition of the vulnerabilities mentioned in the specification file from a local database.
Briefly, the repository is a copy of the online NVD repository (see Section \ref{sec:vulnerability} for more details) where each vulnerability identifier is associated with a statement.
Statements are replaced through a pre-processing step.
The output is an SMT specification (see Section~\ref{sec:semantics}) that we process with the SMT solver CVC4~\cite{Barrett11cvc4}.\footnote{Since SMT specification language is standard, CVC4 can be replaced or put in parallel with any other solver.}
CVC4 returns either \emph{unsat} or a model satisfying the specification.
In the first case, the specification cannot be instantiated.
Unsatisfiability means that the scenario is either contradictory or exceeds the quota of the tenant.\footnote{The two cases can be easily disambiguated by checking the satisfiability of the scenario without quota constraints.}
In both cases, the tenant has useful feedback for refining her specification.

Instead, if a model is generated it is used to feed the infrastructure instantiation process (translator).
The translator converts the definitions of the model into corresponding entries of the Packer and Terraform scripts (see Section~\ref{sec:agen}).
These scripts are the input for infrastructure initialization and updating components.
The packer script is only executed once before the initialization to create the OS images needed for the scenario.
Instead, the Terraform scripts are passed to a scheduler that executes them at the right time.
Every script (except for the first one which initializes the infrastructure) causes a modification of the infrastructure, e.g., by disconnecting or adding a node.
The scheduler launches the scripts following a precise timeline (see Section~\ref{sec:agen}).


\section{Virtual Scenario Description Language}
\label{sec:vsdl}

In this section, we introduce VSDL.
Due to its rich syntax, we only present part of it and we provide the basic intuition through the application to our working example. 

\subsection{VSDL syntax}
\label{sec:syntax}

A VSDL specification describes a scenario in terms of its core elements, i.e., \emph{nodes} and \emph{networks}.
Both of them are defined through a group of statements.
Statements cover a plethora of aspects, e.g., connectivity, firewall rules, and hardware profiles, and they can be composed through standard logic connectives.
Also, a statement can lay under a temporal guard establishing when, during the scenario, it must hold.

\begin{table}[t]
\begin{center}
\caption{Excerpt of the VSDL syntax.}
\label{tab:syntax}
\rule{\columnwidth}{1pt}
\smallskip
\noindent
\begin{small}
\begin{tabular}{r l}
S ::=& \texttt{scenario} Id TTU \texttt{\{} E$^*$ \texttt{\}} \\
TTU ::=& $\varepsilon$ $\:\mid\:$ \texttt{duration} TI \\
TI ::=& Nat \texttt{m} $\:\mid\:$ Nat \texttt{h} \\
E ::=& NODE $\:\mid\:$ NET \\
NODE ::=& \texttt{node} Id \texttt{\{} GNO$^*$ \texttt{\}} \\
GNO ::=& \texttt{[} UN \texttt{] ->} NO $\:\mid\:$ NO \\
NO ::=& NOA $\:\mid\:$ \texttt{not} NO $\:\mid\:$ NO \texttt{and} NO $\:\mid\:$ \\
 & NO \texttt{or} NO $\:\mid\:$ \texttt{(} NO \texttt{)} \\
NOA ::=& \texttt{type is } (\texttt{compute} $\mid \cdots \mid$ \texttt{same as} Id) $\:\mid\:$ \\
 & \texttt{flavour is } (\texttt{mobile} $\mid \cdots \mid$ \texttt{same as} Id) $\:\mid\:$ \\
 & \texttt{cpu is } (\texttt{equal to} Freq $\mid \cdots \mid$ \texttt{same as} Id) $\:\mid\:$ \\
 & \texttt{disk is } (\texttt{equal to} Size $\mid \cdots \mid$ \texttt{same as} Id) $\:\mid\:$ \\
 & \texttt{OS is } (Id $\mid$ \texttt{same as} Id $\mid \cdots $) $\:\mid\:$ \\
 & \texttt{mounts software }Id $\:\mid\:$ \texttt{exists user }Id $\:\mid\:$ \\
 & \texttt{user }Id \texttt{can} (\texttt{read} $\mid$ \texttt{read} $\mid$ \texttt{exec}) Path $\:\mid\:$ \\
 & \texttt{contains } (\texttt{file} $\mid$ \texttt{directory}) Path $\:\mid\:$ \\
 & \texttt{suffers from }Vuln $\:\mid\:$ \\
 & $\cdots$ \\ 
NET ::=& \texttt{network} Id \texttt{\{} GNE$^*$ \texttt{\}} \\
GNE ::=& \texttt{[} UN \texttt{] ->} NE $\:\mid\:$ NE \\
NE ::=& NEA $\:\mid\:$ \texttt{not} NE $\:\mid\:$ NE \texttt{and} NE $\:\mid\:$ \\
 & NE \texttt{or} NE $\:\mid\:$ \texttt{(} NE \texttt{)} \\
NEA ::=& \texttt{bandwidth is} (\texttt{equal to} BW $\mid \!\cdots\! \mid$ \texttt{same as} Id) $\:\mid\:$ \\
 & \texttt{gateway} \texttt{has} \texttt{direct} \texttt{access} \texttt{to} \texttt{the} \texttt{Internet} $\:\mid\:$ \\
 & \texttt{addresses range from} Addr \texttt{to} Addr $\:\mid\:$ \\
 & \texttt{firewall blocks} (\texttt{port} Nat $\mid$ \texttt{IP} Addr) $\:\mid\:$ \\
 & \texttt{firewall} \texttt{forwards} (\texttt{port} Nat \texttt{to} Nat $\mid$ \\
\multicolumn{2}{c}{$\hookrightarrow$ \texttt{IP} Addr \texttt{to} Addr) $\:\mid\:$} \\
 & \texttt{node} Id (\texttt{is connected} $\mid$ \texttt{has IP} Addr) $\:\mid\:$ \\
 & $\cdots$ \\  
UN ::=& UNA $\:\mid\:$ \texttt{not} UN $\:\mid\:$ UN \texttt{and} UN $\:\mid\:$ UN \texttt{or} UN $\:\mid\:$ \texttt{(} UN \texttt{)} \\
UNA ::=& \texttt{switch} (\texttt{on} $\mid$ \texttt{off}) \texttt{at} Id . TExp \\
\end{tabular}
\end{small}
\smallskip
\rule{\columnwidth}{1pt}
\end{center}
\end{table}

The syntax of VSDL is given in Table~\ref{tab:syntax}, where---for the sake of simplicity---we omit some statements and we focus on the most illustrative ones.
A scenario S has a name (Id), a duration (TTU) and a sequence of elements E.\footnote{We use a semicolon to separate the terms of a sequence.}
The duration can be unspecified ($\varepsilon$) or equal to a given interval (TI) in hours or minutes.
An element E can be either a node (NODE) or a network (NET). 
Both nodes and networks have a unique identifier and a sequence of guarded node statements (GNO and GNE). 
The GNO and GNE statements can either have a guard UN or not (unguarded statements NO and NE). 
Guards can be atomic (UNA) or obtained by applying the standard boolean connectives, i.e., negation, conjunction and disjunction.
An atomic guard $\texttt{switch on at}~t . P(t)$ says that the guarded statement becomes true at time $t$.
Also, $t$ must satisfy a predicate $P$ defined through a TExp expression, i.e., a boolean expression on time intervals possibly including other (previously declared) time variables $t',t'',\ldots$ 
The guard $\texttt{switch off at}~t . P(t)$ works symmetrically by stating that the guarded statement becomes false at $t$.

Unguarded statements NO and NE can be atomic statements (NOA and NEA, respectively) or composed ones, i.e., by means of the logical connectives.
Atomic statements for nodes and networks are different in order to capture the respective peculiarities.
Node statements are mostly self-explanatory and include type (compute vs. storage), hardware (CPU speed, disk size, etc.), OS and installed software, users and privileges (read, write, execute), and content of the file system (files and directories).
The only statement that requires more attention is \texttt{suffers from} and we will discuss it in Section \ref{sec:vulnerability}.
Network statements include bandwidth, access to the public network, the range of addresses that can be assigned to connected nodes, firewall rules (e.g., port forwarding and address filtering), and network participants.
The following examples illustrate the use of the VSLD.
\begin{example}
\label{ex:node}
Consider the following node blocks.

\begin{lstlisting}
node Phone {
  flavour is mobile;
  not (disk is larger than 8 GB);
  not (cpu is faster than 2 GHz);
  (OS is Android-21) or (OS is Android-19);
}

node ApacheS {
  flavour is server;
  disk is larger than 200 GB;
  cpu is faster than 8 GHz;
  OS is Debian-8;
  mounts software apache2;
  mounts software php5;
  mounts software dvwa-setup.sh;
}
\end{lstlisting}

Briefly, it contains the statements for nodes ``Phone'' and ``ApacheS''.
The first node represents a smartphone in the scenario and must have an adequate hardware profile, i.e., mobile flavor.
Also, specific hardware constraints can be specified.
For instance, here we force Phone to have 8 GB of disk space at most.
Moreover, we want Phone to have a CPU slower than 2 GHz.
For what concerns the software running on Phone, the requirement is that it mounts Android version 5.0 (API level 21) or 4.4 (API level 19).

ApacheS represents a server hosting an Apache/PHP web application.
Thus, we require the node to have a server flavor, with more than 200 GB of disk and a CPU faster than 8 GHz.
Also, the OS of the server is Debian Linux version 8.
Moreover, we force the server to install three pieces of software: Apache 2.x HTTP server, PHP5, and dvwa-setup.sh.
The last one is a script setting up the Apache server with DVWA~\cite{Dvwa16webpage}.
\end{example} 

\begin{example}
\label{ex:network}
Consider the following VSDL fragment.
\begin{lstlisting}
network Laboratory {
  addresses range from 8.8.8.1 to 8.8.8.64;
  node RSLaptop has IP 8.8.8.3;
  [switch off at t.t < 40 m] -> node Phone is connected;
}

network Main {
  gateway has direct access to the Internet;
  node Laboratory is connected;
  firewall blocks port 22;
  firewall forwards port 80 to 8080;
  firewall blocks IP 8.8.8.1;
}
\end{lstlisting}

The fragment contains two network elements, i.e., Laboratory and Main.
Nodes connected to Laboratory will have IP addresses ranging from 8.8.8.1 to 8.8.8.64.
Both Phone and RSLaptop are connected to Laboratory.
RSLaptop must have the specific address 8.8.8.3.
The last statement states that Phone will leave the sub-network after 40 minutes.

The Main network is connected to the Internet.
Also, Laboratory is a sub-network of Main.
The last three statements define the firewall rules: the firewall must block any incoming connection on port 22, redirect messages using port 80 to port 8080 and prevent any communication with address 8.8.8.1.
\end{example}

\subsection{VSDL semantics}
\label{sec:semantics}

The semantics of VSDL is given through a translation into a SMT~\cite{Barrett09smt} specification.
An SMT specification is a sequence of assertions over the values assumed by functions and constants.
The domains of functions define the \emph{theory} under which the formula must be satisfiable.
VSDL statements refer to several complex data types, e.g., IP addresses, time, node, and network identifiers.
We reduce all of them to the domain of positive numbers.\footnote{For time intervals one might prefer to consider real numbers rather than positive ones (with the time unit set to 1 minute). This would require a change of theory, but this does not affect to rest of the dissertation.} 
All the relevant aspects of the scenario are encoded through one or more dedicated functions, called \emph{description functions}.
Below we list some of the most interesting ones along with their meaning in natural language.

\begin{itemize}
\item node.disk$(t : \mathbb{N}_0, n : \mathbb{N}_0) = s : \mathbb{N}_0$. At time $t$ node (identified by) $n$ has a disk size of $s$ MB.
\item node.cpu$(t : \mathbb{N}_0, n : \mathbb{N}_0) = s : \mathbb{N}_0$. At time $t$ node $n$ has a CPU speed of $s$ MHz.
\item node.app$(t : \mathbb{N}_0, n : \mathbb{N}_0, s : \mathbb{N}_0) = b : \mathbb{B}$. At time $t$ node $n$ mount software $s$ if and only if $b =$ true.
\item node.user.canr$(t : \mathbb{N}_0, n : \mathbb{N}_0, u : \mathbb{N}_0, r : \mathbb{N}_0) = b : \mathbb{B}$. At time $t$, on node $n$, user $u$ can read resource $r$ if and only if $b =$ true.
\item network.gateway.internet$(t : \mathbb{N}_0, n : \mathbb{N}_0) = b : \mathbb{B}$. At time $t$ network $n$ is directly connected to the Internet if and only if $b =$ true.
\item network.node.address$(t : \mathbb{N}_0, n : \mathbb{N}_0, m : \mathbb{N}_0) = a : \mathbb{N}_0$. At time $t$ node $m$ is connected to network $n$ with address $a$ (if $a = 0$ the node $m$ is not connected to $n$).
\item network.firewall.address.forward$(t : \mathbb{N}_0, n : \mathbb{N}_0, a : \mathbb{N}_0) = b : \mathbb{N}_0$. At time $t$ the firewall of network $n$ forwards incoming packets with destination $a$ to $b$ (if $b = 0$ the packets is blocked).
\item network.firewall.port.forward$(t : \mathbb{N}_0, n : \mathbb{N}_0, p : \mathbb{N}_0) = q : \mathbb{N}_0$. At time $t$ the firewall of network $n$ forwards incoming packets on port $p$ to $q$ (if $q = 0$ the packet is blocked).
\end{itemize}

Assertions belong to three groups that we describe below.
\begin{description}
\item[Scenario.] These constraints consist of a direct translation of the VSDL statements into SMT assertions.
Most of the statements have a straightforward interpretation, e.g., if the specification of node $n$ says that, at time $t$, the CPU speed is faster than $800$ MHz, it will result in \texttt{(assert (> (node.cpu t n) 800))}.

\item[Resources.] Some constraints are posed by the computational resources available to the physical infrastructure, i.e., the quota of the tenant.
Such constraints are obtained by translating the OpenStack quota information (see Section~\ref{sec:arch}) into corresponding assertions.
For instance, if the OpenStack quota limits the total CPU frequency to $K$ THz, we must include an assertion stating that the summation of the speed of all the CPUs appearing in the scenario cannot exceed $K$. 

\item[Invariants.] This category includes the constraints that must be constantly satisfied by a scenario and, possibly, are implicitly entailed by the specification. 
For instance, we often want to ensure that all the nodes defined in a scenario are distinct (notice that this might not be always required).
Another assumption might be that nodes belonging to the same network do not share their IP address.
\end{description}

\begin{example}
\label{ex:semantics}
Consider again the specifications given in Examples~\ref{ex:network} and~\ref{ex:node}.
The statements for Lababoratory, Phone and ApacheS are translated into a specification resembling that given in Table~\ref{tab:smt}.

\begin{table}
\caption{Fragment of the SMT specification obtained from Examples~\ref{ex:node} and~\ref{ex:network}.}
\label{tab:smt}
\begin{lstlisting}[
language=lisp,mathescape,morekeywords={forall,sum,declare-fun},
  columns=fullflexible,
  basicstyle=\small,  
  numbers=left,
  numbersep=5pt,
  numberstyle=\color{gray}
]
; Scenario elements
(declare-fun Phone () Int)
(declare-fun ApacheS () Int)
(declare-fun Laboratory () Int)
; ...
; Hardware constraints: Phone
(assert (forall ((u Int)) (and (< (node.cpu u Phone) 16192) (< (node.disk u Phone) 32768))))
(assert (forall ((u Int)) (and (>= (node.cpu u Phone) 512) (>= (node.disk u Phone) 2048))))
(assert (forall ((u Int)) (< (node.disk u Phone) 8192)))
(assert (forall ((u Int)) (< (node.cpu u Phone) 2048)))
; ...
; Network constraints: Laboratory
(assert (forall ((u Int) (n Int)) 
 (or 
  (and 
   (<= (network.node.address u n Laboratory) 134744065) 
   (>= (network.node.address u n Laboratory) 134744128)
  ) 
  (= (network.node.address u n Laboratory) 0)
)))
(assert (forall ((u Int)) (= (network.node.address u RSLaptop Laboratory) 134744067)))
(declare-fun t () Int)
(assert (< t 40))
(assert (forall ((u int)) 
 (and 
  (=> (<= u t) (> (network.node.address u Phone Laboratory) 0))
  (=> (> u t) (not (> (network.node.address u Phone Laboratory) 0))) 
)))
; ...
; Network constraints: Main
(assert (forall ((u Int)) (network.gateway.internet u Main)))
(assert (forall ((u Int)) (> (network.node.address u Laboratory Main) 0)))
(assert (forall ((u Int)) (= (network.firewall.port.forward u Main 22) 0)))
(assert (forall ((u Int)) (= (network.firewall.port.forward u Main 80) 8080)))
(assert (forall ((u Int)) (= (network.firewall.address.forward u Main 134744065) 0)))
; ...
; Invariants
(assert (not (= Phone ApacheS)))
(assert (not (= Phone Laboratory)))
(assert (not (= ApacheS Laboratory)))
; ...
\end{lstlisting}
\end{table}

The meaning of the specification is the following.
Each scenario element, i.e., nodes and networks, declared in the VSDL specification is translated into a corresponding constant (lines 2-4).
Description functions are constrained to assume certain values depending on the VSDL statements.
For instance, the assertions at lines 7-10 encode the statements defining the hardware profile of Phone (see Example~\ref{ex:node}).\footnote{Disk size and CPU speed are given in MB and MHz, respectively.}
Lines 13-28 encode the properties of the Laboratory network.
In particular, any node n must have either an IP address taken from the defined interval (lines 16-17)\footnote{The IP address $a.b.c.d$ is encoded through the formula $d + {2^8}c + {2^{16}}b + {2^{24}}a$.} or the constant $0$ for ``disconnected'' (line 19).
Intuitively, line 21 states that node RSLaptop is connected to a specific address.
Lines 22-28, encoding the conditional statement of Example~\ref{ex:node}, require more attention.
First of all, a constant t is declared (line 22).
Then t is constrained by the \texttt{switch off} guard expression, i.e., t $<$ 40 (line 23).
Similarly to the case for RSLaptop, we translate the statement \texttt{node Phone is connected} by asserting that network.node.address(u, Phone, Laboratory) must be greater than $0$.
This assertion is put in the scope of a double implication (lines 26 and 27) such that the assertion, initially true, must be false after t. 
Lines 31-35 encode the statements for network Main.
Briefly, Main is connected to the Internet (line 30) and Laboratory is connected to Main (line 32).
Also, the firewall of Main blocks (i.e., forwards to 0) packets on port 22 (line 33), forwards packets on port 80 to port 8080 (line 34) and blocks packets directed to address 8.8.8.1.    
Finally, the invariants block states that all the nodes in the specification must be distinct.
\end{example}


\subsection{Vulnerability injection}
\label{sec:vulnerability}

Enabling a vulnerability in existing infrastructures while keeping them realistic is a rather complex task.
As discussed in Section \ref{sec:back}, nowadays a common approach is to run on a certain node a piece of software, e.g., a web application, where many vulnerabilities can be enabled through a proper configuration.
This substantially simplifies the work of the attackers and defenders who only have to discover them. 

In the last years, several actors put considerable effort into compiling vulnerability reports and keeping repositories up to date.
Among them, the National Vulnerability Database\footnote{\url{https://nvd.nist.gov/}} (NVD) represents a major proposal for the standardization of vulnerability reports.
NVD records include a unique id, a textual description, various scores and, more interestingly, a \emph{list of known, vulnerable configurations}.
Each configuration consists of a (often) simple formula, i.e., a disjunction of affected components in CPE (Common Platform Enumeration) format.

Intuitively, a CPE is a unique identifier for hardware/software configuration.
The basic scheme of a CPE is
\begin{center}
\texttt{cpe:/\{prt\}:\{vnd\}:\{prd\}:\{ver\}:\{upd\}:\{edt\}:\{lan\}}
\end{center}
\noindent
where
\begin{itemize}
\item \{prt\} is a single character indicating whether the CPE refers to a class of applications (a), operating systems (o) or hardware (h);
\item \{vnd\} is the vendor of the product;
\item \{prd\} is the product name;
\item \{ver\} is the version number of the product;
\item \{upd\} is the product update identifier;
\item \{edt\} is the product edition name, and;
\item \{lan\} is the product language tag.
\end{itemize}

Also, notice that some fields are optional (meaning \lq\lq any value\rq\rq when omitted) and few wildcards are allowed.

\begin{example}
\label{ex:vulnerability}
The NVD record CVE-2015-0235\footnote{https://web.nvd.nist.gov/view/vuln/detail?vulnId=CVE-2015-0235} reports a heap-based buffer overflow vulnerabilities that permits the remote execution of arbitrary, unauthorized instructions.
The record indicates the following two vulnerable configurations.
\begin{lstlisting}[language=XML, mathescape]
<vuln:vulnerable-configuration>
 <cpe-lang:logical-test operator="OR" negate="false">
  <cpe-lang:fact-ref name="cpe:/a:oracle:communications:13.1"/>
  <cpe-lang:fact-ref name="cpe:/a:oracle:pillar_axiom:6.1"/>
  <cpe-lang:fact-ref name="cpe:/a:oracle:pillar_axiom:6.2"/>
  <cpe-lang:fact-ref name="cpe:/a:oracle:pillar_axiom:6.3"/>
 </cpe-lang:logical-test>
</vuln:vulnerable-configuration>
<vuln:vulnerable-configuration>
 <cpe-lang:logical-test operator="OR" negate="false">
  <cpe-lang:fact-ref name="cpe:/a:gnu:glibc:2.0"/>
  $\ldots$
  <cpe-lang:fact-ref name="cpe:/a:gnu:glibc:2.17"/>
 </cpe-lang:logical-test>
</vuln:vulnerable-configuration>
\end{lstlisting}

The first configuration happens when some specific Oracle software is installed.
In particular, three versions of Pillar Axiom~\cite{oracleaxiom}, i.e., from 3.1 to 3.3, suffer from the vulnerability.
Moreover, a second family of software modules enables CVE-2015-0235.
Such a family includes several versions of the GNU C library glibc.
\end{example}

The translation from the NVD vulnerable configuration to VSDL is quite straightforward and it is obtained through the statement \texttt{suffers from}.
The statement contains a reference to a vulnerability identifier taken from the NVD and its meaning is that the corresponding node must host one of the vulnerable configurations described above.
In practice, it is a shorthand for a composition of \texttt{OS is} and \texttt{mounts software} statements corresponding to the structure of the NVD vulnerable configuration.

\begin{example}
Consider again the vulnerability of Example \ref{ex:vulnerability} and the following VSDL node specification.

\begin{lstlisting}
node N {
  suffers from "CVE-2015-0235";
}
\end{lstlisting}

It is equivalent to the following one.

\begin{lstlisting}[mathescape]
node N {
  (
    mounts software communications-13.1
    or mounts software pillar_axiom-6.1
    or mounts software pillar_axiom-6.2
    or mounts software pillar_axiom-6.3
  )
  or  
  (
    mounts software glibc-2.0
    $\ldots$
    or mounts software glibc-2.1
  );
}
\end{lstlisting}

\end{example}


\section{Automatic Generation of Scenarios}
\label{sec:agen}

The output of the SMT solvers follows a standard syntax~\cite{Barrett15smt}.
In particular, a model consists of a finite sequence of function definitions given through \texttt{define-fun} statements.
Under our assumptions, models contain $(i)$ a list of node identifiers, i.e., a constant for each node and network, $(ii)$ a list of time switches, i.e., constants identifying the instants at which the scenario changes its state, and $(iii)$ a definition for each of the description functions introduced in Section~\ref{sec:semantics}.

It is important to notice that SMT solvers tend to generate minimal models, i.e., those that satisfy the input specification by assigning smaller values to the variables.
In principle, this is a desirable property as it guarantees that our approach generates compact scenarios using a minimal amount of computational resources to satisfy the given constraints.
Pragmatically, scenario designers might find the output model simplistic.
In this respect, the model provides useful information that can be used to refine the original specification.
If the model obtained from scenario $S$ is satisfactory, terraform and packer scripts are generated as follows.
\begin{enumerate}
\item For each time switch $t_i$ (including $t_0 = 0$) a terraform script called \texttt{"$S\_t_i$.tf"} is created. 
All the scripts are initialized with OpenStack access details, including user and tenant name, password, and authorization URL (necessary for establishing a valid session).
\item For each node identifier $n$ $(i)$ a corresponding terraform node resource is added to each script \texttt{"$S\_t_i$.tf"} and $(ii)$ a packer JSON script \texttt{$n$.json} is created.
The latter defines the OS image to be used for the initialization of node $n$.
The former contains the \texttt{image\_name} of the packer and the \texttt{name} field.
\item For each network identifier $m$ a corresponding terraform router resource, together with a suitable interface and port, is added to each script \texttt{"$S\_t_i$.tf"}.
\item For each description function $f$ a group of corresponding terraform commands is added to resource $k$ in script \texttt{"$S\_t_i$.tf"}.
Such commands depend on the value that $f$ assumes on $t_i$ and $k$.
\end{enumerate}

Below we show the outcome of the procedure previously described when applied to our working example.

\begin{example}
\label{ex:terraform}
Consider the SMT specification of Example~\ref{ex:semantics}.
The output generated by an SMT solver invoked over it is similar to that given in Table~\ref{tab:model}.
\begin{table}[t]
\caption{Excerpt of the model generated from the specification of Example~\ref{ex:semantics}.}
\label{tab:model}
\begin{lstlisting}[
language=lisp,mathescape,morekeywords={forall,sum,model,define-fun, ite},
  columns=fullflexible,
  basicstyle=\small,  
  numbers=left,
  numbersep=5pt,
  numberstyle=\color{gray}
]
(model
(define-fun Phone () Int 1)
(define-fun ApacheS () Int 2)
(define-fun RSLaptop () Int 3)
(define-fun Laboratory () Int 4)
(define-fun Main () Int 5)
;...
(define-fun t () Int 1)
;...
(define-fun node.cpu ((p1 Int) (p2 Int)) Int (ite (= p2 1) 512 (ite (= p2 2) 8193 (ite ...)))
(define-fun node.disk ((p1 Int) (p2 Int)) Int (ite (= p2 1) 2048 (ite (= p2 2) 204801 (ite ...)))
(define-fun network.gateway.internet ((p1 Int) (p2 Int)) Bool (ite (= p2 4) false (ite (= p2 5) true (ite ...)))
;...
)
\end{lstlisting}
\end{table}
It consists of a list of functions and constants definitions.
Constants for nodes and networks are assigned to distinct identifiers, i.e., positive numbers (lines 2-6), while time constants are mapped to specific, possibly overlapping minutes of the scenario duration (line 8).\footnote{Notice that t can also be legally assigned to 0, which implies (see Examples~\ref{ex:network} and~\ref{ex:base}) that the Phone node is never connected to (or immediately disconnected from) the Laboratory network. If this is not the intention of the designer, it means that the scenario is underspecified.}

Functions (lines 10-12) are slightly more complex.
They consist of a finite composition of conditional statements (\emph{if-then-else}, \texttt{ite}) testing the value of (some of) the formal parameters of a function to decide the result.
For instance, node.cpu (line 10) is assigned to the partial function $f : \mathbb{N}_0 \times \mathbb{N}_0 \rightharpoonup \mathbb{N}_0$ defined as follows.
\[
f = \lambda\, p1, p2. 
\left\{ \begin{array}{l @{\qquad} l}
512 & \textnormal{if } p2 = 1 \\
8193 & \textnormal{if } p2 = 2 \\
\:\vdots \\
\bot & \textnormal{otherwise} \\
\end{array}
\right.
\]

Assuming no other time constants exist, the model given above results in two terraform scripts \texttt{"S\_0.tf"} and \texttt{"S\_1.tf"}.
The content of the script \texttt{"S\_0.tf"} includes the fragment shown in Table~\ref{tab:script}.

\begin{table}
\caption{Excerpt of the terraform script \texttt{"S\_0.tf"}.}
\label{tab:script}
\begin{lstlisting}[
language=bash,morekeywords={resource, name, network, flavour_name, admin_state_up, external_gateway, network_id, cidr, router_id, subnet_id, fixed_ip, port, flavour_name, image_name},
  columns=fullflexible,
  basicstyle=\small,  
  numbers=left,
  numbersep=5pt,
  numberstyle=\color{gray}
]
# ...
resource "openstack_networking_router_v2" "main" {
  name = "Main"
  external_gateway="b998c866-f909-48a3-a5d6-7837fe91354d"
}

resource "openstack_networking_router_v2" "laboratory" {
  name = "Laboratory"
}
# ...
resource "openstack_networking_subnet_v2" "laboratory" {
  name = "Laboratory"
  network_id = "${openstack_networking_network_v2.laboratory.id}"
  cidr = "8.8.8.1/26"
}
# ...
resource "openstack_networking_router_interface_v2" "laboratory_router" {
  router_id = "${openstack_networking_router_v2.main.id}"
  subnet_id = "${openstack_networking_subnet_v2.laboratory.id}"
}
# ...
resource "openstack_networking_port_v2" "phone_laboratory" {
  network_id = "${openstack_networking_network_v2.laboratory.id}"
  fixed_ip {
    subnet_id = "${openstack_networking_subnet_v2.laboratory.id}"
  }
}

resource "openstack_compute_instance_v2" "phone" {
  name = "Phone"
  image_name = "android-4.4-x86_64"
  flavour_name = "mobile.phone"
  network {
    port = "${openstack_networking_port_v2.phone_laboratory.id}"
  }
}
# ...
\end{lstlisting}
\end{table}

Briefly, it contains the declarations of OpenStack resources for the initialization of the infrastructure.
Among them, routers (lines 2-9) are labeled with ``openstack\_network\-ing\_router\_v2''.
Simply, they include a name attribute identifying them.
Moreover, the router Main has a reference to a (predefined) gateway for accessing the Internet (line 4).
The script binds routers with sub-networks (lines 11-15) through a specific interface (lines 17-20).
Sub-networks also include an address mask in CIDR notation. 
Finally, computational nodes are labeled with ``openstack\_compute\_instance\_v2'' (lines 29-36) and connected to a network through a port (lines 22-27).
The attributes of a node define its name, the OS image to be installed on it, and its hardware profile.\footnote{Currently, Terraform does not support detailed hardware description, e.g., CPU speed. We avoid this issue by dynamically customizing the OpenStack flavours, e.g., ``mobile.phone''.}
Once submitted, the script results in the virtual infrastructure appearing in Figure~\ref{fig:dash}.
Also, as shown in Figure~\ref{fig:dvwa}, DVWA is actually running on the appointed node.
Exploiting a command injection vulnerability, for instance, an attacker can run \texttt{nmap} to check which IP address in the local sub-network correspond to active hosts. 

\begin{figure*}[t]
\begin{center}
\includegraphics[width=\textwidth]{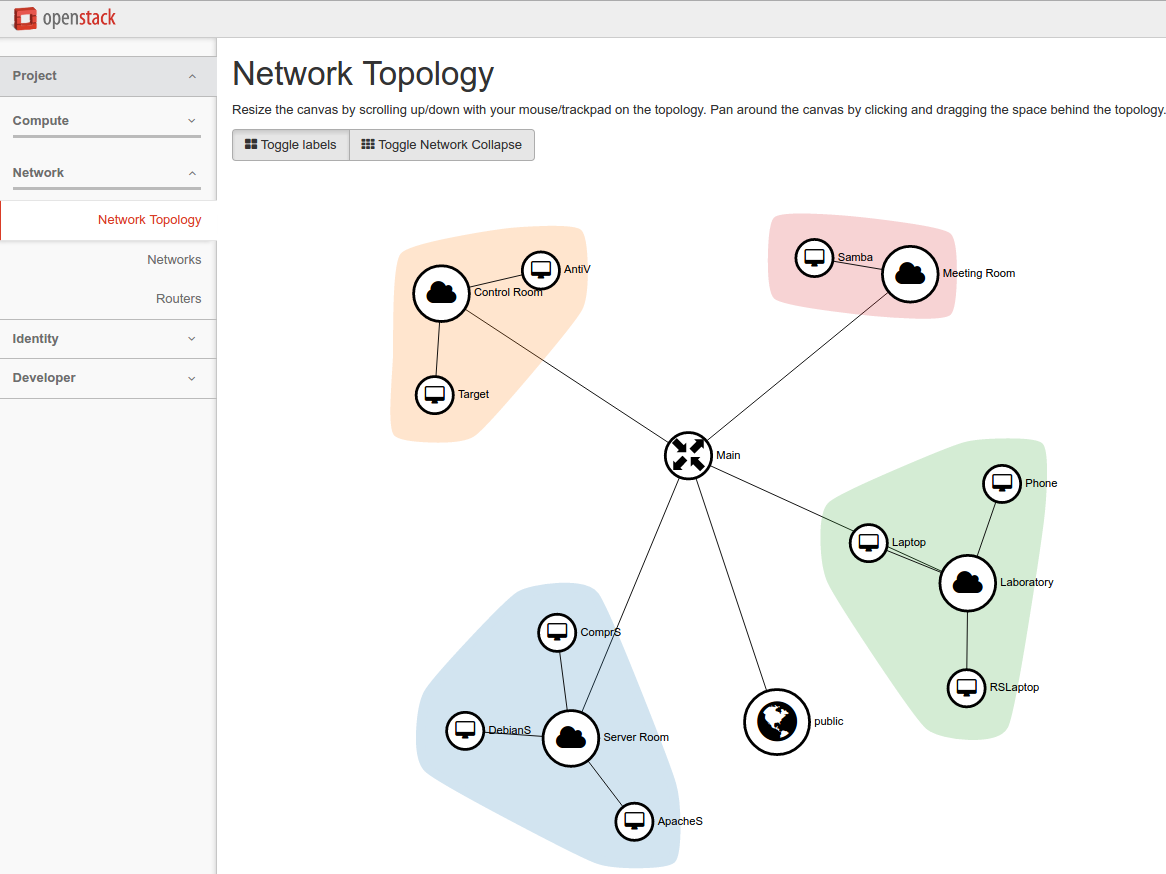}
\caption{The virtual infrastructure as displayed by the OpenStack dashboard.}
\label{fig:dash}
\end{center}
\end{figure*}

\begin{figure*}[t]
\begin{center}
\includegraphics[width=0.8\textwidth]{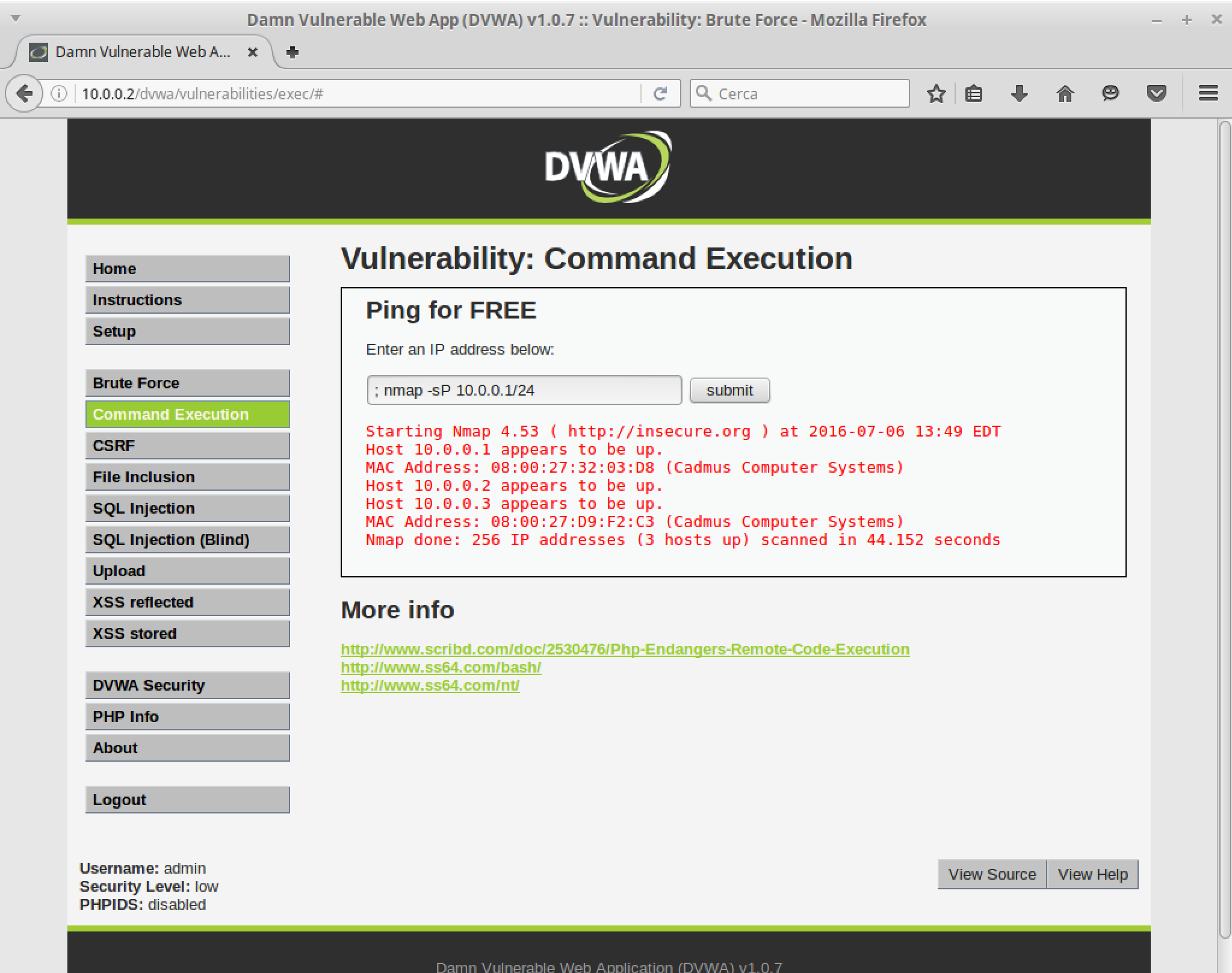}
\caption{The working instance of DVWA exposing a command injection vulnerability.}
\label{fig:dvwa}
\end{center}
\end{figure*}
\end{example}


\section{Discussion}
\label{sec:discussion}

In this section, we critically discuss our proposal's main novelties and limitations.

\begin{description}
\item[Verifiability.] As we highlighted in the previous sections, SMT-based verification allows one to formally check the correctness of the VSDL scenarios.
Nevertheless, other approaches are viable and one might want to consider other languages. 
As a matter of fact, existing SMT theories mostly deal with domains that are not of great interest for developing cyber range scenarios, e.g., think of real numbers.
\item[Expressiveness.] As for the previous case, the high expressiveness of SMT may not always be desirable.
Indeed, less expressive languages could provide better performances and more scalability in large scenarios.
For instance, in~\cite{Russo2020} Datalog is used in place of SMT.
\item[Compositionality.] 
Our high-level DSL supports and simplifies the creation of reusable components and configurations.
Moreover, our verification process allows designers to quickly and automatically validate scenarios so avoiding composition errors. 
This substantially reduces the effort for the creation of training environments.
\item[Integration.]
By relying on a mainstream IaaS provider, our scenarios can be integrated into many contexts and they can take advantage of existing infrastructures.
In general, one might want to achieve integration with other types of infrastructures, e.g., made of physical devices.
For the time being this is not supported by our approach, but technologies such as Software-Defined Networking~\cite{Kreutz2015} could be considered for this aim.
\end{description}
\section{Conclusion}
\label{sec:conc}

We presented a framework for the definition, validation, and generation of virtual scenarios, being at the very core of every cyber range.
To the best of our knowledge, this is the first proposal for such a framework.
Our approach offers several desirable features in terms of verifiability and maintainability.
Moreover, we developed and integrated it with state-of-the-art technologies.
Last but not least, we plan to apply our framework to the forthcoming Italian national cyber range.

This is only the first step toward a fully automated system and many aspects still need to be considered and investigated.
We schematically report those that are the more challenging and relevant.
We account for all of them as future work.

\begin{description}
\item[Traffic simulation.] Although an infrastructure can be detailed from an architectural point of view, the scenarios based on it might lack of realism in terms of network traffic.
Traffic generators exist, e.g., \emph{Ostinato}\footnote{\url{http://ostinato.org/}}, but they need to be configured to correctly simulate the real behavior of the infrastructure.
This will also require extending the syntax of VSDL with statements for describing the network activity of a node.
\item[Infrastructure inference.] For the time being, virtual scenarios are designed by experts to resemble a real infrastructure.
Obtaining a VSDL model from monitoring/observing an actual infrastructure would lead to an easier and more realistic modeling process.
For instance, the output of tools like \emph{EtherApe}\footnote{\url{http://etherape.sourceforge.net/}} and \emph{Xplico}\footnote{\url{http://www.xplico.org/}} could be (partially) translated to VSDL.
\item[Attack trees.] Vulnerabilities play a central role, in particular, in the training sessions.
Although we can effectively inject vulnerabilities process, forcing their exploitation through predefined steps following a didactic purpose is not trivial.
We plan to extend our framework with predefined sets of attack trees that a scenario designer can include in her VSDL specification.
Since attack trees include sub-goals, we must make sure that the scenario permits the exploration of the tree and includes at least one attack pattern. 
\item[Infrastructure fuzzing.] As discussed in Section~\ref{sec:agen}, our framework generates a minimal infrastructure satisfying the specification.
Nevertheless, it is a common practice to have rich scenarios with myriads of nodes having no specific/active roles in the attack/defense process.
To this aim, we aim at including fuzzing methodologies that can add complexity to a scenario without compromising its key features.
\end{description}


\label{sect:bib}
\bibliographystyle{unsrt}
\bibliography{main}
------------------------------------------------------------------------------
\section*{Author Biography}
\vspace*{1em}
\begin{biography}{Gabriele Costa}{photo/gc} is Associate Professor at the SySMA Group of the IMT School for Advanced Studies. He received his M.Sc. in Computer Science in 2007 and his Ph.D. in Computer Science in 2011. He was a member of the cybersecurity group of the Istituto di Informatica e Telematica (IIT) of the CNR. His appointments include a period as visiting researcher at ETH Zurich in 2016-2017. He has been co-founder of the Computer Security Laboratory (CSec) at DIBRIS. He is co-founder and CRO of Talos, a spin-off of DIBRIS focused on Cybersecurity. His focus is on studying and applying formal methods for the automatic verification and testing of mobile and modular systems.
\end{biography}
\vspace*{0.5em}
\begin{biography}{Enrico Russo}{photo/er}  is Assistant Professor in Computer Engineering and received his M.Sc. in Computer Science in 2001 and his Ph.D. in Computer Science in 2021. His work is focused on Cyber Range systems with particular emphasis on investigating techniques for simplifying the generation of the training environments and for automatizing the tasks of personnel involved in exercises. His research interests also include maritime cyber security. In 2019, he co-founded ZenHack, the Capture the Flag team of the University of Genova, and is personally involved in different live-fire cyber exercises as a Green/Blue team member.
\end{biography}
\vspace*{4em}
\begin{biography}{Alessandro Armando}{photo/aa} is Full Professor in Computer Engineering and received his Ph.D. in Computer Engineering at the University of Genoa. His appointments include a position as research fellow at the University of Edinburgh and one at INRIA-Lorraine (France). He is professor at the University of Genoa where he teaches Computer Security. He has contributed to founding and directing the Master in Cybersecurity and Data Protection of the University of Genoa and the Master in Digital Forensic and Cyber Technologies at the School of Telecommunications of the Defence. He is currently serving as director of the Ph.D. Program in Security, Risk and Vulnerability of the University of Genoa. He founded and led the Security \& Trust Research Unit of the Bruno Kessler Foundation in Trento. He has been coordinator and/or team leader in several national and EU research projects, including a European Industrial Doctorate in partnership with SAP. He contributed to the discovery of authentication flaws in the Single Sign-On standards and implementations, including a serious man-in-the-middle attack on the SAML-based SSO for Google Apps. He is currently serving as deputy director of the Cybersecurity National Laboratory of the National Inter-university Consortium on Informatics (CINI).
\end{biography}
\end{document}